**Research Article**

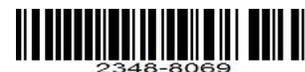

# Growth and proline content in NaCl stressed plants of annual medic species

## Chérifi Khalil[1*], Haddioui Abdelmajid[2], El Hansali Mohammed[2], Boufous El Houssein[3]

[1] Laboratory of Biotechnology and Valorization of Natural Resources, Faculty of sciences, Ibn Zohr University, P. O. Box 8106, 80000 Agadir, Morocco

[2] Laboratory of management and Valorization of Natural Resources, Genetic and Phytobiotechnology, Faculty of sciences and technics, Sultan Moulay Slimane University, B.P. 523, Beni Mellal, Morocco

[3] Department of Biochemistry and Microbiology, Laval University, Quebec city (Quebec), Canada

*Corresponding author: kcherifi@yahoo.com

## Abstract

Wild populations of *Medicago ciliaris* and *Medicago polymorpha* were subjected to four salt treatments 0, 50, 100 and 150 mM NaCl, plant growth and proline concentration in leaves were assessed. The analyzed data revealed significant variability in salt response within and between the two species, depending on the salinity level. It was found that high NaCl concentrations affected all the growth parameters. However, the reduction was more important at higher NaCl concentrations and the highest reduction was obtained for the populations of *Medicago polymorpha* where it reached around 90% in root length at 150 mM NaCl for Pmar. The Tunisian population of *Medicago ciliaris*, prospected on soils affected by salinity, was the best tolerant in all ecotypes studied in this work. This population, exhibits a particular adaptability to salt environment at both germination and seedling stage. Furthermore, the correlation among the studied plants sensitivity and leaf proline concentration showed that high proline contents were related to their reactivity to salt. Consequently, it appeared that proline biosynthesis occurred presumably as a consequence of disturbance in cell homoeostasis and reflected poor performance and greater damage in response to salt stress. These findings indicated that this osmolytes content may be used as another useful criterion to differentiate salt-tolerant from salt sensitive plant in annual medics.

## Introduction

Salinity is one of the major abiotic stresses affecting plant production in arid and semi-arid regions. Salt stress induces both osmotic stress and ionic stress(Ueda et al., 2004). The osmotic stress is triggered by low osmotic potential which limiting absorption of water from soil, and the ionic stress is caused by the over-accumulation of toxic salt ions within plant cells. These stresses, as well as the nutritional imbalance affect the physiological status and are responsible for deleterious effects to plant growth and development(Santos et al., 2001; Meringer et al., 2016). To cope with these unfavorable environmental conditions, plants have developed different protective mechanisms to preserve normal cellular metabolism and prevent injury to cellular constituents including the accumulation of ions and osmolytes such as proline (Gratão et al., 2015; Kolupaev et al., 2016). Furthermore, salt tolerant plants involved less Na+ and Cl- transport to leaves and was able to maintain lower $Na^+/K^+$ proportions in their shoots and developed higher facility to sequester ions into cytoplasm or cell wall(Lee et al., 2003). The compartmentalization of ions in vacuoles was also more efficient in salt-tolerant than salt-sensitive plants (Radi et al., 2013).

Several plants accumulate higher level of proline in contrast to other amino acids when exposed to drought or a high salt content in the soil (Heidari et al., 2011). Proline has been found to may act as a mediator of osmotic adjustment stabilizing the effect of salt accumulated in the vacuole (Heidari et al., 2011), to protect cell membranes, several different enzymes and metabolic machinery (Zadehbagheri et al., 2014).

In the present study, we focused on two annual species of *Medicago* chosen for their forage quality and their capacity to improve nitrogenation of salted land: *Medicago ciliaris* (L.) and *Medicago polymorpha* (L.). These two diploid (2n=16) autogamous species grow naturally in a semi-arid to arid superior bioclimatic stages in Morocco (Chérifi, 2016). These species, classed as glycophytes, may be observed on relatively saline soils in association with halophytes (Abdelly et al., 2006;Ferchichi et al., 2015). In these regions, the halophytic species do not represent a good pastoral resource, at least for livestock because they are loaded with salt, which accounts for 15 to 30% of dry matter. On the other hand, medics contain ten times fewer NaCl and can produce, in rainy years, up to 40% of the vegetative cover. (Abdelly et al., 1995). Annual species of *Medicago* are very appreciated and supported by the halophytes, which contribute periodically to the preservation of low saline soil. Nevertheless, their productivity can be reduced by 40% in salt concentration to 12 g/l (Chérifi et al., 2011). The use of these species for rehabilitation and exploitation of lands affected by salinity can contribute to install, between the halophytes, an interesting potential of grassland for exploiting this ecosystem like pasturing area.

The aim of the present study was to assess salt tolerance variability at growth seedling stage of *M. ciliaris* and *M. polymorpha*, until now not estimated, in order to explore opportunities for selection and breeding salt tolerant genotypes. The study also investigated the relationship between the intensity of salt tolerance and the accumulation of proline in these *Medicago* species, the level of proline after salt stress was examined.

# Materials and Methods

## Plant materials

Seven wild populations prospected in the South-west of Morocco and one Tunisian population were studied: three populations of *M. ciliaris* and four populations of *M. polymorpha*. For each species, pods representative of each population were collected randomly on diverse regions reputed to be more or less affected by salinity. The four ecotypes of *M. polymorpha* was originated from the areas of Marrakech (Pmar), Taroudant (Ptar), Chtouka Ait Baha (Pchk) and Massa (Pmas). Those of *M. ciliaris* are collected near Marrakech (Cmar), Tétouan (Ctét) and Tunisia (Ctun). This last population was collected near sabkha of Boufiicha city (Chérifi, 2016) and chosen in this work as control population for salt tolerance estimation.

## Seedling cultivation and salinity treatments

To investigate salt tolerance ability, seeds from different pods (for limiting parental effect) were manually scarified and sterilized with 0.5% sodium hypochlorite solution (NaCl) for 10 minutes, then rinsed with sterile distilled water several times, and briefly blotted on filter paper. 30 seeds from each accession were placed in plastic Petri dishes (90 mm diameter) on filter paper wetted with distilled water. After five days, pre-germinated seeds were transplanted into each pot filled with red mountain sand, already sifted. This type of sand has proved to be interesting during preliminary experiments sinceit showed good performance in plant growth and also allow efficient drainage of water.

Seedlings were grown under the same conditions in a plastic tank and watered with distilled water. When the second trifoliate leaf was emerging, pots were irrigated with four salinity levels. To avoid salinity shock and to acclimatize seedlings to high NaCl concentrations, salinity stress was imposed by applying the saline irrigation water progressively. Each treatment was irrigated with the lowest NaCl concentration then with the next higher concentration until each treatment reached its considered irrigation concentration (Al-Khaliel, 2010). The NaCl concentrations were 0, 50, 100 and 150 mM.

## Growth parameters

Twenty plants for each treatment were randomly sampled to determine leaf area and the length of both shoot and root as well as of the longer branch. In addition the total plant dry weight were measured after desiccation in a forced draft oven at 70 C for 48h. For normalizing the results, data were expressed as the relative reduction in comparison to the control using the following formula:

Relative reduction (%) =[(1−(salinized/control)]×100

## Determination of proline content

Proline content was measured by the modified method of (Bates et al., 1973). Seedlings were cultivated under the same conditions as those used in the growth experiment above. Samples (50 mg fresh weight) from the eighth leaf were each extracted with 2 ml of 40% methanol. 1 ml extract was mixed with 1 ml of a mixture of glacial acetic acid and orthophosphoric acid (6 M) (3: 2, v/v) and 25 mg ninhydrin. After one hour of incubation at 100 °C, the tubes were cooled and 5 ml toluene were added. The absorbance of the upper phase was spectrophotometrically determined at 528 nm. The proline concentration was determined using a standard curve and expressed as $\mu g \ g^{-1}$ tissue dry weight. In this experience, the maximum salinity was limited to 100mM NaCl because the high concentration induced premature senescence of the older leaves.

## Statistical analysis

The experiment was set up in a completely randomized design with 20 replications per population and by treatment for growth parameters. For proline dosage, 5 repetitions per population were randomly determined.

All values expressed as a percentage were arcsine square root transformed before performing statistical analysis to normalize the data and to homogenize the variance (Chamorro et al., 2015). All data obtained were subjected to a two-way analysis of variance (ANOVA) with populations and salinity treatments as factors, followed by a Student–Newman–Keuls post hoc test. A difference was considered to be statistically significant when P < 0.05. All statistical analysis were performed with Statistica software version 6.1 for Windows (StatSoft, 2001).

# Results and Discussion

## Effect of NaCl on growth parameters

For all the studied parameters, the variance analysis applied to the data revealed a very high significant population and salinity effect (p < 0,001) Table 1 and 2.

Table 1.F statistics values from the two-way variance analysis (ANOVA) for each growth parameter investigated by population and salinity for *M. ciliaris*.

| Source | Salinity (S) | Population (P) | SxP |
|---|---|---|---|
| Leaf area | 81,088*** | 11,623*** | 0,641 [ns] |
| Shoot length | 42,501*** | 14,649*** | 0,316 [NS] |
| Branch length | 43,363*** | 25,708*** | 0,658 [NS] |
| Root length | 41,147*** | 14,952*** | 1,248 [NSs] |
| Plant dry weight | 6,836** | 15,390*** | 0,219 [NS] |

*\*\*\*: very high significant test; [NS]: non-significant test*

Table 2. F statistics values from the two-way variance analysis (ANOVA) for each growth parameter investigated by population and salinity for *M. polymorpha*.

| Source | Salinity (S) | Population (P) | SxP |
|---|---|---|---|
| Leaf area | 173,475*** | 13,112*** | 2,024 [NS] |
| Shoot length | 146,541*** | 62,219*** | 0,539 [NS] |
| Branch length | 129,226*** | 45,911*** | 6,056*** |
| Root length | 142,937*** | 41,450*** | 4,999*** |
| Plant dry weight | 33,519*** | 10,291*** | 0,520 [NS] |

*\*\*\*: very high significant test; [NS]: non-significant test*

The interactions between the two effects were not significant for the most parameters except branch and root length for *M. polymorpha*. Indeed, for the tow species, NaCl produced a remarkable reduction for all of the considered growth parameters (Fig. 1).

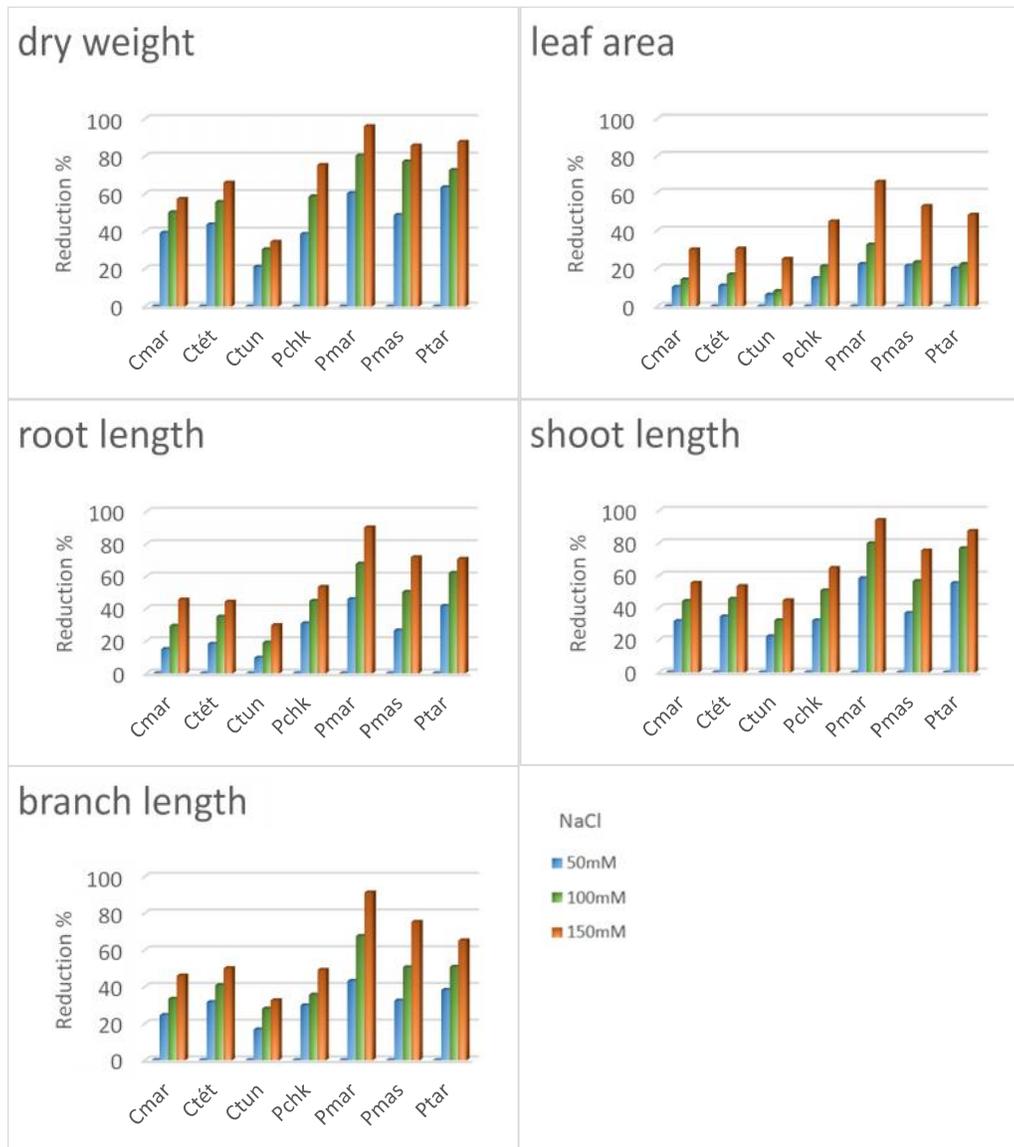

Figure 1. Effect of the salinity on the growth parameters of the populations of *M. ciliaris* an *M. polymorpha*

Nevertheless, the sensitivity of the different parameters varied, the low salinity treatment (50 mM) reduced these parameters to a lesser degree than high salinity treatment. However, the reduction was more important at higher NaCl concentrations. The percentage reduction in root length was greater at 100 and 150 mM NaCl and the highest reduction was obtained for the populations of *Medicago polymorpha,* where it reached around 90% at 150 mM NaCl for Pmar, followed by Pmas and Ptar with approximately 70%. The decrease in this parameter for *Medicago ciliaris* did not exceed 45% at 150 mM NaCl and the lowest value was 29% for Ctun population. Salinity also caused reduction in the total plant dry weight with a greater reduction as the NaCl concentration was augmented. It remained highest also in population Pmar of *M. polymorpha* and reached 96% at 150 mM. The same decrease was much more pronounced in shoot and longer branch length, it was maximum in Pmar population and reached 94 and 91% of the control at 150 mM, respectively, for shoot and branch length. For the percentage reduction in leaf area, it appeared that this parameter was the least affected but it varied mostly in the same way for all populations of the two species.

This finding was in agreement with the large genotypic variability in salt response revealed also at germination stage within the same ecotypes (Chérifi et al., 2011). However, interspecific differences were noted based on the growth measured parameters. The *M. polymorpha* populations showed greater sensitivity at higher salinities, especially Pmar population when *M. ciliaris* was the least affected ones, particularly at the Tunisian population Ctun originating from salt-affected areas (Sabkha)which was well adapted to saline conditions and therefore was less affected by salinity. This population exhibited also the best tolerance to NaCl at germination stage (Chérifi et al., 2011).

Similar results were reported for other leguminous species, where growth parameters like leaf area, dry weight were significantly reduced with increasing salinity (Ashraf and Iram, 2005;Okçu et al., 2005;Husen et al., 2016). According to several studies, the decrease in growth may either be regarded as a consequence of the requirement of water resulting in insufficient osmotic solutes to generate turgor or to the toxicity of $Na^+$ and/or $Cl^-$ in the plant tissues (Boonsaner and Hawker, 2010).Moreover, the reduction in dry weight under salinity stress may be accredited to inhibition of hydrolysis of reserved foods and their translocation to the growing shoots (Singla and Garg, 2005). This suggests, that salinity stress imposed additional energy supplies on plant cells and reduced photosynthesis, which in turn limited the supply of carbohydrate required for growth process (Cheeseman, 1988). Moreover, salt stress could also reduce root, shoot and leaves expansion by reducing turgor pressure in plant and consequently reducing cell growth and development (Sirousmehr et al., 2014).

Based on these measured growth parameters, salt-stressed plants revealed no characteristic symptoms, and only comparison with normal plants determine the degree of salt stress (Yokota, 2003).

## Proline accumulation

A two-way ANOVA indicated a significant main effect of salinity (P<0.001) and population (P<0.01) on the proline content in plant leaves (Table 3 and 4). However, the salinity x population interaction, which indicate a differential effect of salinity between populations, was no significant in the two species, reflecting that all ecotypes responded similarly to the different NaCl concentrations.

**Table 3.**Results of two-way variance analysis (ANOVA) forproline accumulation by population and salinity for *M. ciliaris*.

| Source | df | MS | F |
|---|---|---|---|
| Salinity (S) | 2 | 6229.34 | 14.29590*** |
| Population (P) | 2 | 2033.038 | 4.66568** |
| S x P | 4 | 852.922 | 1.95740[NS] |

*df: degree of freedom; MS: mean square; \*\*\*: very high significant test; \*\*: high significant test; [NS]: non-significant test*

**Table 4.**Results of two-way variance analysis (ANOVA) for proline accumulation by population and salinity for *M. polymorpha*.

| Source | df | MS | F |
|---|---|---|---|
| Salinity (S) | 2 | 21652.02 | 47.60080*** |
| Population (P) | 3 | 3033.30 | 6.66854** |
| S x P | 6 | 844.23 | 1.85599[NS] |

*df: degree of freedom; MS: mean square; \*\*\*: very high significant test; \*\*: high significant test; [NS]: non-significant test*

The effect of salinity on the proline accumulation in the leaf tissue was examined in order to investigate whether proline accumulation contributes to the adaptation of plants in salt stress. In both species, proline amount in the leaf tissue was found to increase under salinity in all populations of the two species (fig. 2).

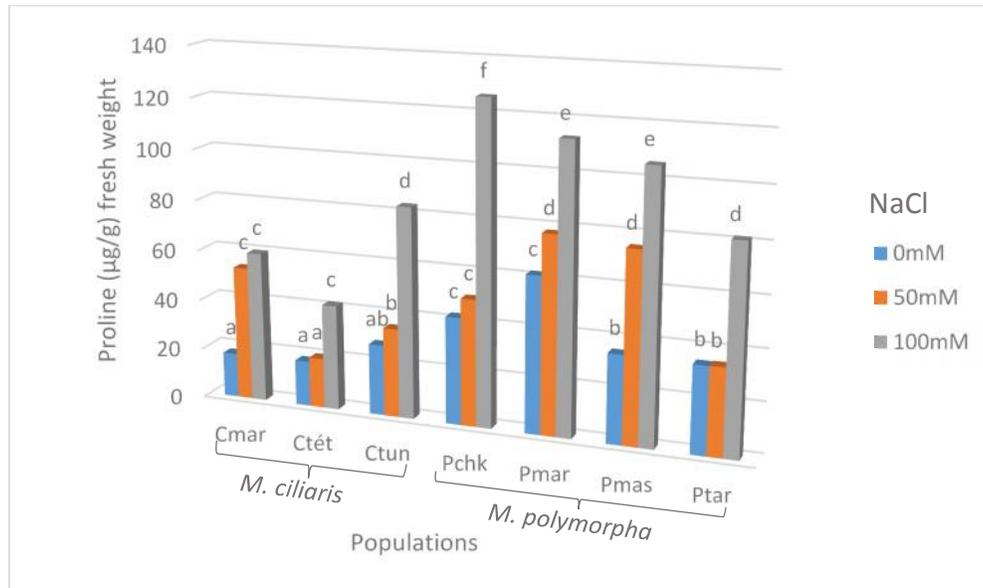

Figure 2. Effect of salinity on the accumulation of proline in the leaves of the populations of
*M. ciliaris* an *M. polymorpha*
*Means of populations having the same letter are not significantly different (p < 0.05) (Newman-Keuls test).*

However, the *M. ciliaris* populations accumulated less proline than those of *M. polymorpha*. At the lower concentration of NaCl the level of proline was only just above that of the control, mostly at Ctét, Ctun, Pmar and Ptar. The greatest accumulation was observed when the plants were exposed to 100 mM NaCl. At this concentration, the Pchk and Ptar showed the highest proline content and the increase was respectively four and two times higher than in control. Generally, the correlation among the studied plants sensitivity, estimated on the basis of growth and leaf proline concentration showed that high proline contents were related to their reactivity to salt.

These results are consistent with the finding of a negative relationship between salt tolerance and proline accumulation (Fougere et al., 1991;Torabi and Halim, 2010;Li et al., 2016;Zheng et al., 2016). The accumulation of proline in the two species could contribute to the cellular adaptation to salt stress or be a result of metabolic changes induced by salinity (Khedr et al., 2003). Consequently, it appeared that proline biosynthesis occurred presumably as a consequence of disturbance in cell homoeostasis and/or of an increase in the use of photosynthesis products for proline biosynthesis at the expense of plant growth (Silambarasan and Natarajan, 2014). Others authors, suggested that proline is a symptom of salt stress injury rather than an indicator of the resistance (Lutts et al., 1999;Jacobs et al., 2003), and reflected poor performance and greater damage in response to salt stress (Moradi and Ismail, 2007).

In contrast, several studies have argued that proline might be thought to play a primordial role to maintain growth when plants were placed in saline condition (Hoson and Wada, 1980). Therefore, proline contributes to osmotic adjustment, stabilization and protection of membranes integrity and macromolecules from the damaging effects of salinity and as hydroxyl radical scavenger (Molinari et al., 2007;Liu et al., 2013).Furthermore, it has been proposed that proline accumulation in leaves occurred to preserve chlorophyll content and turgor to protect the photosynthetic activity under salt stress conditions (Yıldıztugay et al., 2011).

## Conclusions

In the light of the above result, it is concluded that:

On the basis of growth parameters, the results showed variability in salt response within and between *M. ciliaris* and *M. polymorpha* genotypes, depending on the salinity level.

Salt stress decreased growth significantly, but sensitivity differed for both species: The highest reduction was obtained for the populations of *Medicago polymorpha*, where it reached around 90% in root length at 150 mM NaCl for Pmar (the most sensitive).

The Tunisian population of *Medicago ciliaris,* prospected on soils affected by salinity, was the best tolerant in all ecotypes studied in this work. This population, exhibited a particular adaptability to salt environment at both germination and seedling stage. Therefore, this ecotype could be recommended for rehabilitation of damaged salted land.

The intraspecific variation in salt tolerance, revealed at both germination and growth seedling stage, may be used to select genotypes particularly suitable to start breeding programs for salt tolerance.

The results from this study suggested that proline accumulation could be considered as a useful biochemical marker for assessing increased salt stress tolerance in annual *Medicago* species. Thus, it appeared that proline biosynthesis occurred presumably as a consequence of disturbance in cell homoeostasis and reflected poor performance and greater damage in response to salt stress.

Further detailed researches, including additional ecotypes of medic species, are necessary on field conditions. This will provide an opportunity to conduct this investigation to verify correlation between salt tolerance during seed germination and other stages of plant development which could be useful in breeding programs for selecting salt tolerant genotypes in annual *Medicago* species.